\documentclass[11pt,a4paper]{article}\usepackage{jheppub}
\usepackage{amsmath}
\usepackage[most]{tcolorbox}
\usepackage{dsfont}
\usepackage{ulem}
\usepackage{natbib}
\usepackage{xcolor}
\usepackage[hang,flushmargin]{footmisc}
\usepackage{tikz-cd}
\usepackage{enumitem}
\usepackage{amsfonts,amssymb, amscd,amsmath,latexsym,amsbsy,bm}
\usepackage{stmaryrd}
\usepackage{todonotes}
\usepackage{stackrel}
\usepackage{amsthm}

\usepackage{graphicx}

\usepackage{physics}

\title{The Gross-Pitaevskii equation from the Heisenberg spin chain}

\author{Osman Erkan Kaluc$^{a}$, and Ilmar Gahramanov$^{a,b,c}$}
\affiliation{
	$^a$ {Department of Physics, Bogazici University, 34342 Bebek, Istanbul, Turkey}\\[-0.5cm]
	
	$^{b}$ Institute of Radiation Problems, Azerbaijan National Academy of Sciences, \\ B.Vahabzade St. 9, AZ1143, Baku, Azerbaijan\\[-0.5cm]
	
	$^{c}$ Department of Mathematics, Khazar University,  Mehseti St. 41, AZ1096, Baku, Azerbaijan \\[-0.5cm]
}
\emailAdd{erkan.kaluc@boun.edu.tr}
\emailAdd{ilmar.gahramanov@boun.edu.tr}

\abstract{In this paper, we revise the derivation of the Gross-Pitaevskii equation from the Heisenberg spin chain stressing physical aspects and all technicalities.}

\keywords{Gross-Pitaevskii equation, Heisenberg spin chain, coherent states, classical limit, Holstein-Primakoff transformation}

\begin{document}
	\maketitle
	\flushbottom

\section{Introduction}

The Gross-Pitaevskii equation, commonly referred as the nonlinear Schrödinger equation, was derived by Gross in \cite{gross1961structure} and Pitaevskii in \cite{pitaevskii1961vortex} (see, also \cite{Pitaevskii2016}) to describe the ground state behavior of a bosonic system at low temperatures, which exhibit superfluidity features, for a full discussion see e.g. \cite{Tannoudji}. As a nonlinear partial differential equation, it is a remarkable fact that a special case of the Gross-Pitaevskii equation (namely, the one without the potential term) is a classically integrable system and has soliton solutions \cite{shabat1972exact,Zakharov:1974zf,novikov1984theory}. This equation has many physical applications and is a subject of an active research, see, e.g. \cite{copie2020physics,serkin2000novel,kartashov2011solitons,Pazarci:2022rdw}. The most general derivation of the Gross-Pitaevskii equation is done by the application of the second quantization, mean-field approximation of the interaction, and variational approach.

On the other hand, a different derivation of the Gross-Pitaevskii equation (in fact, its color-generalized version, which exhibits $U(p,q)$-symmetry) from the Heisenberg spin chain was suggested in \cite{Dubrovin1988} based on the approach in \cite{MakhankovPashaev1983}. The main purpose of this paper is to provide a detailed treatment of the colorless Gross-Pitaevskii equation, using the coherent states of Heinsenberg-Weyl group. Another alternative approach may be conducted by the use of the coherent states of the $SU(1,1)$ group \cite{pashaev1986nonlinear}.

The interesting problem that we have not touched on, is a possible generalization of the approach followed here to the supersymmetric coherent states. This, in turn, would result in a system of supersymmetric Gross-Pitaevskii equations which exhibit $osp(2|1)$-symmetry, see, e.g. \cite{Makhankov:1988jp}. This system was suggested by Kulish in \cite{Kulish:1985qv} (see, also \cite{Popowicz:1994aw}).

The paper is organized as follows: In Sec. 2 and 3 we review the basic background, including the Heisenberg spin chain, Holstein-Primakoff and Jordan-Wigner transformations. In Sec. 4 we give an explicit derivation of the Gross-Pitaevskii equation. In Appendices, we discuss fermionic coherent states in terms of Grassmann variables and the Hubbard model.

\section{XXZ Heisenberg spin chain}

One of the most important quantum integrable systems is the Heisenberg spin chain model \cite{Heisenberg:1928mqa} which occurs in many areas of mathematical physics \cite{Takhtajan:1979iv,Kirillov:1987zz,Faddeev:1996iy}. In this model, spins are arranged in a one-dimensional lattice with the lattice constant $c$. We will consider the model with the following Hamiltonian 
\begin{align}
 \hat{H}&=\hat{H}_s+\hat{H}_L \;\text{,}
\end{align}
where $\hat{H}_L$ represents the harmonic lattice Hamiltonian given by
\begin{align}
 \hat{H}_L=\frac{m}{2}\sum_j x^2_j+\frac{m v^2_0}{2c^2}\sum_j(x_{j+1}-x_j-c)^2 \;\text{.}
\end{align}
Here $j$ is indexing the spin site. The second term $\hat{H}_s$ is the XXZ Heisenberg Hamiltonian in a non-homogeneous external field directed along the $(Oz)$-axis given by
\begin{align}
 \hat{H}_S=-\frac{1}{2}\sum_{i,\sigma}\left[\frac{1}{2}J_{j+\sigma j}\left(S^{+}_{j+\sigma}S^{-}_j+S^{-}_{j+\sigma}S^{+}_j\right)+R_{j+\sigma j}S^{z}_{j+\sigma}S^{z}_{j}\right]-\sum_j h^z_j S^z_j \;\text{,}
\end{align}
where $h_j$ denotes the field strength at the site $j$ and $\sigma=\pm 1$. Note that the case with a homogeneous external field was considered in \cite{makhankov1987generalized}.

\section{Holstein-Primakoff and Jordan-Wigner Transformations}

We will use the Holstein-Primakoff transformation to write the Heisenberg spin chain in terms of the bosonic creation and annihilation operators. For that purpose, let us briefly discuss the meaning of this transformation. We know from the $\mathfrak{su}(2)$ algebra that the spin states can be expressed in terms of two numbers $j$ and $m$, which give the eigenstate $\ket{j,m}$ of the complete set of commuting operators $S^2$ and $S_z$ with respect to the Hilbert space of the spin degrees of freedom. The idea of the Holstein-Primakoff transformation starts by introducing the so-called "spin-deviation" operator defined as
\begin{align}
\mathfrak{n}_z=S-S_z \;\text{.}
\end{align}
Observe that for each fixed value of $S$, to each $j$ corresponds a unique eigenvalue $\mathfrak{n}_z$ \cite{HolPrim1}. Hence, we can write $\ket{\mathfrak{n}_z}$ instead of $\ket{j,m}$. The spin deviation measures the deviation of the $z$-component of the spin from its maximum value. Upon writing explicitly the eigenvalue equations of the spin operators $S^{+}$, $S^{-} $, and $S^{z}$, we observe that they obey the same kind of relations as the bosonic creation and annihilation operators. This comparison, taken one step further, leads to suggesting the following transformation formula
\begin{align}
S^{+}_j&=\sqrt{2s}\left(1-\frac{\hat{n}_j}{2s}\right)^{\frac{1}{2}}\hat{a}_j\label{HP1} \;\text{,}\\
S^{-}_j&=\sqrt{2s}\left(1-\frac{\hat{n}_j}{2s}\right)^{\frac{1}{2}}\hat{a}^{\dagger}_j\label{HP2}\;\text{,}\\
S^{z}_{j}&=s-\hat{n}_j\;\text{,}
\end{align}
where $\hat{n}_{j}=a^{\dagger}_{j}a_{j}$ denotes the number operator of the $j$th site. In some sense, we pass to the second-quantization form of the spin operators. The corresponding creation and annihilation operators can be called the Holstein-Primakoff bosons, which indicate the spin-excitations \cite{HolPrim1,HolPrim2,HolPrim3}.

For large values of $s$, we can approximate the first two equations (\ref{HP1})-(\ref{HP2}) as
\begin{align}
S^{+}_j \simeq \sqrt{2s}\hat{a}_j\;\text{,}\\
S^{-}_j \simeq \sqrt{2s} \hat{a}^{\dagger}_j\;\text{.}
\end{align}
The insertion of the expressions results in the following form of the XXZ Hamiltonian
\begin{align}
H_S=-\frac{1}{2}\sum_{\sigma,j}\left[J_{j+{\sigma},j}s(\hat{a}_{j+\sigma}\hat{a}^{\dagger}_j+\hat{a}^{\dagger}_{j+\sigma}\hat{a}_{j})+R_{j+{\sigma},j}(s-\hat{n}_{j+\sigma})(s-\hat{n}_{j})\right] \;\text{.}
\end{align}
Similarly, we have different kinds of transformations such as the transformation from the spin operators to the fermionic creation and annihilation operators, called the Jordan-Wigner transformation. In the case of spin 1/2, the down and up eigenstates can be thought of as the empty and filled states of a fermionic gas. This is the main motivation of Jordan and Wigner. The necessity to satisfy the required commutation relations is solved by the following transformation \cite{coleman_2015,JordanWigner}.
\begin{align}
S^{+}_j&=a^{\dagger}_j \exp(-i\pi \sum^{j-1}_{l=1}a^{\dagger}_la_l) \;\text{,}\\
S^{-}_j&= \exp(i\pi \sum^{j-1}_{l=1}a^{\dagger}_la_l)a_j\;\text{,}\\
S^{z}_j&=a^{\dagger}_ja_j-\frac{1}{2} \;\text{.}
\end{align}
We can show the following results \cite{coleman_2015}
\begin{align}
S^{+}_j S^{-}_{j+\sigma}=a^{\dagger}_ja_{j+\sigma} \;\text{,}\\
S^{-}_j S^{+}_{j+\sigma}=a_{j+\sigma}a^{\dagger}_j\;\text{.}
\end{align}
Thanks to these, the use of Jordan-Wigner transformation leads to the same expression in terms of creation and annihilation operators, but this time exactly.

\section{Equation of Motion and the Coherent States}
We will derive the Gross-Pitaevskii equation starting from the Heisenberg spin chain. Our approach follows and discusses the approach in \cite{Dubrovin1988, MakhankovPashaev1983,MakhankovMyrzakulov1987}, based on the methods developed in \cite{Fedyanin1977}.

\subsection{Gross-Pitaevskii Equation in Bosonic Case \label{GPbosesection}} 
In order to obtain the equations of motion of the operator $a_i$, we pass to the Heisenberg picture and use the Heisenberg equation given by 
\begin{align}
 i \frac{d A_H}{dt} =-[H_H; A_H]\;\text{.}
\end{align}
First of all, we observe that
\begin{align}
[H_H, a_H]&=H_H a_H -a_H H_H\nonumber\;\text{,}\\
 &=U^{\dagger}HUU^{\dagger}aU-U^{\dagger}aUU^{\dagger}HU\nonumber\;\text{,}\\
 &=U^{\dagger}[H,a]U\;\text{.}
\end{align}
Now we need to calculate the commutator $[H_s, a_i]$ in the Heisenberg picture. As the Hamiltonian $H_s$ is made of creation and annihilation operators, we insert $U^{\dagger}U$ between each of them. Therefore, the required commutator in the Heisenberg picture is obtained by replacing all $a_i$ and $a^{\dagger}_i$ in the commutator expressed in the Schrödinger picture with $a_H$ and $a^{\dagger}_H$. For that reason, from now on, we omit the $H$ subscript and unless explicitly mentioned otherwise, all the operators will be taken in the Heisenberg picture. 

Recall that the bosonic creation and annihilation operators (ignoring the spin degrees of freedom) satisfy
\begin{align}
 [a_i, a^{\dagger}_j]&=\delta_{ij} \;\text{,}\nonumber \\
 [a_i, a_j]&=[a^{\dagger}_i, a^{\dagger}_j]=0\;\text{,}
\end{align}
together with the identity operator, this is just the Heisenberg-Weyl algebra.

We need to calculate the following commutator
\begin{align} \nonumber
[H_{s},a_i]&= \frac{1}{2}sJ_{i+1,i}a_{i+1}+\frac{1}{2}sJ_{i-1,i}a_{i-1}+\frac{1}{2}sJ_{i,i+1}a_{i+1}+\frac{1}{2}sJ_{i,i-1}a_{i-1}\\ \nonumber
&-\frac{1}{2}(R_{i,i+1}+R_{i,i-1})sa_i-\frac{1}{2}(R_{i+1,i}+R_{i-1,i})sa_i+\frac{1}{2}(R_{i+1,i}a^{\dagger}_{i+1}a_{i+1}a_i+R_{i-1,i}a^{\dagger}_{i-1}a_{i-1}a_i)\\
&+\frac{1}{2}(R_{i,i+1}a_{i+1}a^{\dagger}_{i+1}a_i+R_{i,i-1}a_{i-1}a^{\dagger}_{i-1}a_i)-h^z_ia_i \;.
\end{align}

Using the Heisenberg equation, we get the equation of motion of the annihilation operator \footnote{If we used the fermionic anticommutation relations which we recall in the Appendix after conducting the Jordan-Wigner transformation, then we get the same answer. This is an expected result as the equation we obtain, the Gross-Pitaevskii equation, is non-relativistic. Thus, the form of the equation does not distinguish fermions and bosons in an intrinsic manner. The main difference will come only from our treatment of complex and Grassmanina numbers.}
\begin{align}
 -i \Dot{a_i}&=\frac{1}{2}sJ_{i+1,i}a_{i+1}+\frac{1}{2}sJ_{i-1,i}a_{i-1}+\frac{1}{2}sJ_{i,i+1}a_{i+1}+\frac{1}{2}sJ_{i,i-1}a_{i-1}\nonumber\\
 &-\frac{1}{2}(R_{i,i+1}+R_{i,i-1})sa_i-\frac{1}{2}(R_{i+1,i}+R_{i-1,i})sa_i+\frac{1}{2}(R_{i+1,i}a^{\dagger}_{i+1}a_{i+1}a_i+R_{i-1,i}a^{\dagger}_{i-1}a_{i-1}a_i)\nonumber\\
 &+\frac{1}{2}(R_{i,i+1}a_{i+1}a^{\dagger}_{i+1}a_i+R_{i,i-1}a_{i-1}a^{\dagger}_{i-1}a_i)-h^z_ia_i \;\text{.} \label{annihiloperatoreq}
\end{align}

\subsection{Coherent state representation}

Now, we will pass to the representation we deem to call coherent state representation. The first appearance of the coherent states is due to the solution of Schrödinger in \cite{Schrodinger1926}. Yet, the modern concept is firstly developed by Glauber in \cite{Glauber1963} where he suggests the use of the coherent states to describe the coherent beam of light. Even if a first attempt to generalize the notion of coherent states in a group-theoretic approach is carried by Barut and Girardello in \cite{Barut1971}, this was limited to non-compact groups. In 1972, Perelomov in \cite{Perelomov1972} and Gilmore in \cite{Gilmore1972} gave a more general definition of the coherent states attached to a Lie group, which today we refer as Perelomov-Gilmore coherent states. According to this definition, a coherent state set can be obtained by the application of the unitary irreducible representation of the Lie group on a fixed state in the Hilbert space, therefore it is an orbit of a chosen fixed state under a given unitary irreducible representation. The choice of the fixed state has the consequence that the coherent states have minimal uncertainty. Moreover, they will form an overcomplete set of states and they are non-orthogonal. It is a fundamental property of the coherent states that they lead to a partition of identity, making possible the expansion of arbitrary states and operators in terms of coherent states \cite{Perelomov1972,Perelomov1977,Perelomovbook,Smorodinskii1992}.

The coherent state representation has therefore the advantage of being the most classical state of the quantum mechanical system. They can therefore be used in the transition from quantum case to the classical case \cite{Smorodinskii1992,Perelomov1977}. In other words, coherent states allow a certain classical representation of the quantum equations. The use of the coherent states and the further manipulation hereupon can be seen as a transition from the quantum integrability of some spin chain models to the classical Liouville integrability of the corresponding nonlinear partial differential equations.

Our derivations are based on the algebra of the bosons, which is the same as the Heisenberg-Weyl algebra, by the very nature of the second quantization. This algebra has exactly the same coherent states, introduced by Glauber. We also discuss the fermionic coherent states and the analogous results. As a different case, a derivation of the classical equations of motion for the coherent states associated with $SU(2)$ is conducted in \cite{MakhankovMyrzakulov1987}.

In the case of the Heisenberg-Weyl group, recall that the coherent states turn out to be the eigenstates of the annihilation operator
\begin{align}
 a\ket{\phi}=\phi\ket{\phi} \;,
\end{align}
where $\phi$ is an arbitrary complex number.

On the other hand, there exists a particularly important property of coherent states:
If the Hamiltonian of the quantum system can be expressed as a linear combination of the generators of the Lie algebra of the group, then the evolution operator is an element of the group itself. It then turns out that a coherent state will always remain coherent during the evolution under this Hamiltonian \cite{Smorodinskii1992}. In this case, the expected value of the annihilation operator is found to be
\begin{align}
\bra{\phi}a_H(t)\ket{\phi}&= \bra{\phi}U(t)^{\dagger}a(t)U(t)\ket{\phi}\nonumber\\
&=\bra{\phi(t)}\phi(t)\ket{\phi(t)}\nonumber\\
&=\phi(t) \; \text{,}
\end{align}
which is a complex-valued function of a real variable.
We now turn to the case of the operators. Whenever we can write operators in their normal (Wick) form
\begin{align}
 \hat{A}=\sum_{m,n} C_{mn}(a^{\dagger}_j)^m(a_j)^n \; \text{,}
\end{align}
the average of $A$ can be easily computed
\begin{align}
A&:=\bra{\phi_i}\hat{A}\ket{\phi_i}\nonumber\\
&= \sum_{mn} C_{mn} (\phi^{*}_i)^m(\phi_i)^n \; \text{,}
\end{align}
which corresponds to the so-called symbol of the operator by Berezin \cite{Berezin1971}.

In our case, we have different spins sitting at different sites $j$. Therefore, the complete state of the system is given by the following tensor product
\begin{align}
\ket{\phi}=\bigotimes_j \ket{\phi_j} \; \text{.}
\end{align}
Hence upon applying $\bra{\phi}--\ket{\phi}$ on (\ref{annihiloperatoreq}), we obtain the following result
\begin{align} \nonumber
-i\hbar\Dot{\phi_i}=&\frac{1}{2}sJ_{i+1,i}\phi_{i+1}+\frac{1}{2}sJ_{i-1,i}\phi_{i-1}+\frac{1}{2}sJ_{i,i+1}\phi_{i+1}+\frac{1}{2}sJ_{i,i-1}\phi_{i-1}\\ \nonumber
 &-\frac{1}{2}(R_{i,i+1}+R_{i,i-1})s\phi_i-\frac{1}{2}(R_{i+1,i}+R_{i-1,i})s\phi_i+\frac{1}{2}R_{i+1,i}|\phi_{i+1}|^2\phi_i+\frac{1}{2}R_{i-1,i}|\phi_{i-1}|^2\phi_i\\ \nonumber
 &+\frac{1}{2}R_{i,i+1}|\phi_{i+1}|^2\phi_i+\frac{1}{2}R_{i,i-1}|\phi_{i-1}|^2\phi_i-h^z_i\phi_i\\
 &=sJ_{i+1,i}\phi_{i+1}+sJ_{i,i-1}\phi_{i-1}-R_{i+1,i}s\phi_i-R_{i-1,i}s\phi_i+R_{i+1,i}|\phi_{i+1}|^2\phi_i+R_{i-1,i}|\phi_{i-1}|^2\phi_i-h^z_i\phi_i \label{eqmotannih}\;,
\end{align}
where we assume the isotropy of the interaction to the sum of the same power terms with coefficients $R_{ij}$ and $R_{ji}$, $J_{ij}$ and $J_{ji}$.

We now pass to the continuum limit where the lattice constant $c$ converges to zero. Then, the coupling constants can be approximated as follows
\begin{align}
 J_{jj}&:=J(0) \;\text{,}\\
 J_{j,j+\sigma} &\simeq J(0)-J(1)|x_j-x_{j+\sigma}|\;\text{,}\\
 R_{jj}&:=R(0)\;\text{,}\\
 R_{j,j+\sigma}& \simeq R(0)-R(1)|x_j-x_{j+\sigma}|\;\text{.}
\end{align}
Moreover, the eigenvalues $\phi_i$ will then become
\begin{align}
 \phi_j &\to \phi(\xi)\;\text{,}\\
 \phi_{j\pm 1} & \simeq \phi(\xi)\pm \frac{d}{d \xi}\phi(\xi)+\frac{1}{2}\frac{d^2}{d\xi^2}\phi(\xi)\;\text{,}
\end{align}
The complete result after eliminating terms of order higher than two is
\begin{align}
 i\hbar \Dot{\phi(\xi)}&=(-2J(0)+2R(0))s\phi(\xi)-sJ(0)\phi_{\xi \xi}(\xi)+2J(1)sx_{\xi}\phi(\xi)-2R(1)sx_{\xi}\phi(\xi)-2R(0)|\phi(\xi)|^2 \phi(\xi)\nonumber\\
 &+2R(1)x_{\xi}|\phi(\xi)|^2\phi(\xi)-2R(0)|\phi_\xi|^2\phi(\xi)-R(0)\phi^{*}(\xi)\phi_{\xi \xi}-R(0)\phi(\xi)\phi^{*}_{\xi \xi}-h^z(\xi)\phi(\xi)\;\text{.}
\end{align}

Note that we have additional dispersive terms 
\begin{align}
 2R(1)x_{\xi}|\phi(\xi)|^2\phi(\xi)-2R(0)|\phi_\xi|^2\phi(\xi)-R(0)\phi^{*}(\xi)\phi_{\xi \xi}-R(0)\phi(\xi)\phi^{*}_{\xi \xi}\;.
\end{align}
In order to eliminate them, we will make a transformation both in the dependent and independent variables, which we choose to be linear for simplicity. The transformations we use are
\begin{align}
 \phi &\to A \varphi \;\text{,} \\
 t &\to \frac{1}{2A^2 R(0)}t' \;\text{,}\\
 \xi &\to B\xi' \;\text{,}
\end{align}
where the coefficients are taken as follows
\begin{align}
 A&=\left(\frac{(-2J(0)+2R(0)+2J(1)x_{\xi}-2R(1)x_{\xi})s}{2R(0)}\right)^{\frac{1}{2}} \;\text{,}\\
 B&=\left(\frac{sJ(0)}{(2R(0)+2J(1)x_{\xi}-2R(1)x_{\xi}-2J(0))s}\right)^{\frac{1}{2}}\;\text{.}
\end{align}
After carrying out the transformation, we obtain
\begin{align}
i\Dot{\varphi}=&\varphi(\xi')-\varphi_{\xi' \xi'}(\xi')-|\varphi(\xi')|^2 \varphi(\xi')+\frac{R(1)B^{-1}x_{\xi'}|\varphi(\xi')|^2\varphi(\xi')}{R(0)}-B^{-2}|\varphi_\xi'|^2\varphi(\xi')\nonumber\\
&-\frac{B^{-2}\varphi^{*}(\xi')\varphi_{\xi' \xi'}}{2A}-\frac{B^{-2}\varphi(\xi')\varphi^{*}_{\xi' \xi'}}{2A}-\frac{1}{2|A|^2R(0)}h^z(\xi)\phi(\xi)\;\text{.} \label{gpprecursor}
\end{align}
Upon requiring
\begin{align}
2 R(0)<< sJ(0) \;\text{,}
\end{align}
we get the Gross-Pitaevskii equation in the presence of a non-homogeneous external field, and where the interaction potential is chosen to be the contact potential, namely,
\begin{align}
i\Dot{\varphi}=&\varphi(\xi')-\varphi_{\xi' \xi'}(\xi')-|\varphi(\xi')|^2 \varphi(\xi')-V(\xi')\varphi(\xi') \;\text{,}
\end{align}
where we noted
\begin{align}
 V(\xi'):=\frac{1}{2|A|^2R(0)}h^z(\xi')\;\text{.}
\end{align}
Note that the passage to the classical integrability is conducted in three steps: passage to the "coherent-state representation", continuum limit, and an assumption on the interaction constants and spin number $s$.

\section{Appendix}
\subsection{Grasmannian Case of the Fermionic Coherent States}

As mentioned in the footnote of section \ref{GPbosesection}, we get the same equation of motion for the fermionic annihilation operator $a_i$. In order to pass to the coherent state representation, we have two possible choices. The first choice is to choose the coherent states of the Heisenberg-Weyl group, which in this case, corresponds to the group generated by the momentum and position operators of the fermionic field. In this particular choice, we will get the exact same equation as (\ref{gpprecursor}), which eventually leads to the ordinary Gross-Pitaevskii equation of a complex-valued function of a real variable.

Another choice is to consider the coherent states of the fermions. Let us recall the fermionic anticommutation relations
\begin{align}
 \{a_{i}, a^{\dagger}_{j} \}&=a_i a^{\dagger}_j+a^{\dagger}_ja_i=\delta_{ij}\;\text{,} \\
 \{a_{i}, a_{j}\}&=a_{i}a_{j}+a_{j}a_{i}=0\;\text{,}\\
 \{a^{\dagger}_{i}, a^{\dagger}_{j}\}&=a^{\dagger}_{i}a^{\dagger}_{j}+a^{\dagger}_{j}a^{\dagger}_{i}=0\;\text{,}
\end{align}
where we ignore the spin degrees of freedom.

In literature, there are two ways to define the coherent states of fermionic systems. One way is to construct a Lie algebra out of these operators. This question has many solutions \cite{Gilmore1983}. The most natural way is to consider the $\mathfrak{spin}(2n)$ algebra \cite{Perelomovbook}. Another method passes through the introduction of the Grassmann numbers. The two approaches can be proven to be equivalent \cite{combescure_2012}. We carry our calculations using the Grassmanians.

In this formalism, we define the fermionic coherent states following \cite{CahillGlauber1999} as follows
\begin{align}
 \ket{\theta}=e^{-\theta a^{\dagger}}\ket{0}\; \text{,}
\end{align}
indexed by a Grassmann number $\theta$, which satisfies the relation $\theta_i \theta_j=-\theta_j\theta_i$.

It is an easy fact to establish that the fermionic coherent states are the eigenstates of the fermionic annihilation operator
\begin{align}
 a \ket{\theta}=\theta\ket{\theta} \;\text{.}
\end{align}
The above relation analogous to the bosonic case makes all the further manipulations exactly the same as in the subsection \ref{GPbosesection}. Therefore, we will obtain the following equation upon applying $\bra{\theta}--\ket{\theta}$ on (\ref{annihiloperatoreq})
\begin{align} \nonumber
 -i\hbar\Dot{\theta_i}=sJ_{i+1,i}\theta_{i+1}+sJ_{i,i-1}\theta_{i-1}-R_{i+1,i}s\theta_i-R_{i-1,i}s\theta_i+R_{i+1,i}|\theta_{i+1}|^2\theta_i+R_{i-1,i}|\theta_{i-1}|^2\theta_i-h^z_i\theta_i \;,
\end{align}
and we take the same kind of continuum limit. Yet, this time the equation we get contains a Grassmannian valued function of a real variable $\theta(\xi)$.

\subsection{Hubbard Model}
A similar pattern gives rise to the coupled Gross-Pitaevskii equation if we start with the Hubbard model
\begin{align}
H_{Hubbard}&=-t\sum_{\sigma,j}\sum_{s}(\hat{a}_{j+\sigma,s}\hat{a}^{\dagger}_{j,s}+\hat{a}^{\dagger}_{j+\sigma,s}\hat{a}_{j,s})+\sum_{j}U_j n_{j,1}n_{j,0} \nonumber \; \text{,}\\
 &=\underbrace{-t\sum_{\sigma,j}\sum_{s}(\hat{a}_{j+\sigma,s}\hat{a}^{\dagger}_{j,s}+\hat{a}^{\dagger}_{j+\sigma,s}\hat{a}_{j,s})}_\textrm{$H_{hub1}$}+\underbrace{\sum_{j}U_j\hat{a}^{\dagger}_{j,1} \hat{a}_{j,1}\hat{a}^{\dagger}_{j,0}\hat{a}_{j,0}}_\textrm{$H_{hub2}$} \; \text{,}
\end{align}
where $s$ denotes the following number
\begin{align}
 s=S_z+\frac{1}{2} \; \text{.}
\end{align}

Our derivation follows the work \cite{MakhankovPashaev1983}. To start, let us recall the fermionic anticommutation relations for fermions considering their spin is then given by
\begin{align}
 \{a_{i,s}, a^{\dagger}_{j,s'} \}&=a_i a^{\dagger}_j+a^{\dagger}_ja_i=\delta_{ij}\delta_{ss'}\; \text{,} \\
 \{a_{i,s}, a_{j,s'}\}&=a_{i,s}a_{j,s'}+a_{j,s'}a_{i,s}=0\; \text{,}\\
 \{a^{\dagger}_{i,s}, a^{\dagger}_{j,s'}\}&=a^{\dagger}_{i,s}a^{\dagger}_{j,s'}+a^{\dagger}_{j,s'}a^{\dagger}_{i,s}=0\; \text{.}
\end{align}
As before, we will use the Heisenberg equation to find the equation of motion of the operator $a_{i,\kappa}$. For this purpose, we take the following commutators.
\begin{align}
[H_{hub1},a_{i,\kappa}]&= -t\left( \left\{\sum_s \sum_{\sigma=\pm 1,j}\left(\hat{a}^{\dagger}_{j,s}\hat{a}_{j+\sigma, s}+\hat{a}^{\dagger}_{j+\sigma,s}\hat{a}_{j,s}\right)\right\} \hat{a}_{i,\kappa}-\hat{a}_{i,\kappa}\left\{\sum_{\sigma=\pm 1,j}\left(\hat{a}^{\dagger}_{j,s}\hat{a}_{j+\sigma,s}+\hat{a}^{\dagger}_{j+\sigma,s}\hat{a}_{j,s}\right)\right\} \right)\nonumber\\
&=t\left\{\sum_s\sum_{\sigma=\pm 1,j}\left(\delta_{ij}\delta_{s,\kappa}a_{j+\sigma,s}+\delta_{i,j+\sigma}\delta_{s,\kappa}a_{j,s}\right)\right\}\nonumber\\
&=2ta_{i+1,\kappa}+2ta_{i-1,\kappa} \; \text{,}
\end{align}
and
\begin{align}
[H_{hub2},a_{i,\kappa}]&= \sum_{j}U_{j}\left(a^{\dagger}_{j,1}a_{j,1}a^{\dagger}_{j,0}a_{j,0}a_{i,\kappa}-a_{i,\kappa}a^{\dagger}_{j,1}a_{j,1}a^{\dagger}_{j,0}a_{j,0}\right)\nonumber\\
&=\sum_{j}U_j(a^{\dagger}_{j,1}a_{j,1}a^{\dagger}_{j,0}a_{j,0}a_{i,\kappa}-\delta_{ij}\delta_{1\kappa}a_{j,1}a^{\dagger}_{j,0}a_{j,0}-\delta_{ij}\delta_{0\kappa}a^{\dagger}_{j,1}a_{j,1}a_{j,0}+a^{\dagger}_{j,1}a_{j,1}a^{\dagger}_{j,0}a_{i,\kappa}a_{j,0})\nonumber\\
&=-U_i\delta_{1\kappa}a_{j,1}a^{\dagger}_{j,0}a_{j,0}-U_i\delta_{0\kappa}a^{\dagger}_{j,1}a_{j,1}a_{j,0}\nonumber\\
&=\begin{cases}
-U_ia_{j,1}a^{\dagger}_{j,0}a_{j,0} \qquad \kappa=1 \\
-U_ia^{\dagger}_{j,1}a_{j,1}a_{j,0}  \qquad \kappa=0 \; \text{.}
\end{cases}
\end{align}

Inserting these into the Heisenberg equation, we then obtain \footnote{We would obtain the exact same results upon using the bosonic commutation relations similar to the section \ref{GPbosesection}}
\begin{align}
-i\hbar \Dot{a}_{i,\kappa}=2t(a_{i+1,\kappa}+a_{i-1,\kappa})-U_ia_{j,1}a^{\dagger}_{j,0}a_{j,0} \qquad \kappa=1 \; \text{,} \label{GPHubbardsystem1}
\end{align}
or
\begin{align}
-i\hbar \Dot{a}_{i,\kappa}=2t(a_{i+1,\kappa}+a_{i-1,\kappa})-U_ia^{\dagger}_{j,1}a_{j,1}a_{j,0}  \qquad \kappa=0 \; \text{.}\label{GPHubbardsystem2}
\end{align}
To pass to the coherent state representation, we have always two possible choices. We can use the coherent states of the Heisenberg-Weyl group or the fermionic coherent states as done in section \ref{GPbosesection}. For the purpose of illustration, we will use the coherent states of the Heisenberg-Weyl group. In case we used the Grassmann number-tagged coherent states as above, note that we would get the same system of equations for Grassmann valued functions.

Upon obtaining the symbol of both sides of (\ref{GPHubbardsystem1}) and (\ref{GPHubbardsystem2}), i.e. applying $\bra{\phi}--\ket{\phi}$, we obtain the following system of equations
\begin{align}
-i\hbar \Dot{\phi}_{i,\kappa}=2t(\phi_{i+1,\kappa}+\phi_{i-1,\kappa})-U_i|\phi_{i,1-\kappa}|^2\phi_{i}\; \text{,} 
\end{align}
We can then pass to the continuum limit, resulting in the following system of equations
\begin{align}
 i\hbar\Dot{\phi_{\kappa}(\xi)}=-2t\left(\phi_{\kappa}(\xi)+\frac{d}{d\xi} \phi_{\kappa}(\xi)+\frac{1}{2}\frac{d^2}{d\xi^2}\phi_{\kappa}(\xi)+\phi_{\kappa}(\xi)-\frac{d}{d\xi} \phi_{\kappa}(\xi)+\frac{1}{2}\frac{d^2}{d\xi^2}\phi_{\kappa}(\xi)\right)+U(\xi)|\phi_{1-\kappa}(\xi)|^2\phi_{\kappa}(\xi)\; \text{,}
\end{align}
which gives upon simplification a system of coupled Gross-Pitaevskii equations
\begin{align}
 i\hbar\Dot{\phi}_{\kappa}(\xi)=-4t\phi_{\kappa}(\xi)-2t\frac{d^2}{d\xi^2}\phi_{\kappa}(\xi)+U(\xi)|\phi_{1-\kappa}(\xi)|^2\phi_{\kappa}(\xi), \qquad \kappa = 0,1 .
\end{align}

\bibliographystyle{utphys}
\bibliography{HeisenbergtoGP}

\end{document}